\documentclass[lettersize,journal]{IEEEtran}
\usepackage{amsmath,amsfonts}
\usepackage{algorithmic}
\usepackage{algorithm}
\usepackage{array}
\usepackage[caption=false,font=scriptsize,labelfont=sf,textfont=sf]{subfig}
\usepackage{textcomp}
\usepackage{stfloats}
\usepackage{url}
\usepackage{verbatim}
\usepackage{graphicx}
\usepackage{amssymb}
\usepackage{cite}
\usepackage{booktabs}
\usepackage{subfloat}
\usepackage{float}
\usepackage{multirow}
\usepackage{soul}
\usepackage{color}

\begin{document}

\title{DASECount: Domain-Agnostic Sample-Efficient Wireless Indoor Crowd Counting\\ via Few-shot Learning}

\author{Huawei Hou,
	    Suzhi Bi,
	    Lili Zheng,
	    Xiaohui Lin,
	    Yuan Wu,
	    and Zhi Quan
        % <-this % stops a space
\thanks{H. Hou, S. Bi, L. Zheng, X. Lin, and Z. Quan are with the College of Electronics and Information Engineering, Shenzhen University, Shenzhen, China 518060 (e-mail: 2070436152@email.szu.edu.cn, \{bsz,zhengll,xhlin,zquan\}@szu.edu.cn). S.~Bi and Z. Quan are also with the Peng Cheng Laboratory, Shenzhen, China 518066. (Corresponding Author: Suzhi Bi)}
\thanks{Y. Wu is with The State Key Lab of Internet of Things for Smart City, and also with the Department of Computer and Information Science, The University of Macau, Taipa, Macao SAR, China (e-mail: yuanwu@um.edu.mo).}

}



\maketitle

\begin{abstract}
Accurate indoor crowd counting (ICC) is a key enabler to many smart home/office applications. 
Recent development of WiFi-based ICC technology relies on detecting the variation of wireless channel state information (CSI) caused by human motions and has gained increasing popularity due to its low hardware cost, reliability under all lighting conditions, and privacy preservation in sensing data processing. 
To attain high estimation accuracy, existing WiFi-based ICC methods often require a large amount of labeled CSI training data samples for each application domain, i.e., a particular WiFi transceiver or background deployment. 
This makes large-scale deployment of WiFi-based ICC technology across dissimilar domains extremely difficult and costly. 
In this paper, we propose a Domain-Agnostic and Sample-Efficient wireless indoor crowd Counting (DASECount) framework that suffices to attain robust cross-domain detection accuracy given very limited data samples in new domains. 
DASECount leverages the wisdom of few-shot learning (FSL) paradigm consisting of two major stages: source domain meta training and target domain meta testing. 
Specifically, in the meta-training stage, we design and train two separate convolutional neural network (CNN) modules on the source domain dataset to fully capture the implicit amplitude and phase features of CSI measurements related to human activities. 
A subsequent knowledge distillation procedure is designed to iteratively update the CNN parameters for better generalization performance.  
In the meta-testing stage, we use the partial CNN modules to extract low-dimension features out of the high-dimension input target domain CSI data. 
With the obtained low-dimension CSI features, we can even use very few shots of target domain data samples (e.g., 5-shot samples) to train a lightweight logistic regression (LR) classifier, and attain very high cross-domain ICC accuracy. 
Experiment results show that the proposed DASECount method achieves over 92.68\%, and on average 96.37\% detection accuracy in a 0-8 people counting task under various domain setups, which significantly outperforms the other representative benchmark methods considered.

\end{abstract}


\section{Introduction}

\IEEEPARstart{A}{utomatic} indoor crowd counting (ICC) has important applications in a number of areas, such as public health management, security monitoring, and home/office automation. 
For instance, in the recent global outbreak of COVID-19, ICC helps maintain social distancing in the indoor environment for effective epidemic prevention. 
Besides, knowing the exact number of people enables to fine-tune the air-conditioner for energy conservation and improve comfort in the indoor office environment.
Existing ICC methods are mainly based on surveillance cameras, wearable sensors and radar, etc \cite{teixeira2010survey}. 
Among them, using cameras raises concerns on privacy violations and is highly susceptible to weak light conditions. 
On the other hand, ICC based on wearable sensors causes additional hardware overhead, e.g., a target needs to wear a special bracelet, which is costly and inconvenient for public or large-scale application scenarios. 
Although ICC based on radar equipment enjoys high detection accuracy when the radars are fined-tuned and properly deployed, the installation and hardware costs are uneconomic for extensive deployment in budget-limited home/office applications. 

In recent years, there has been a growing interest in exploiting WiFi signals for indoor wireless sensing applications. By capturing the impact of human activity on the channel state information (CSI) between the WiFi transmitter and receiver, many indoor wireless sensing tasks can be effectively performed, such as human presence detection, activity and gesture recognition, respiration monitoring, as well as the focus of this paper, indoor crowd counting \cite{r01,r02,r03,r04}. 
Compared with the above-mentioned ICC methods, WiFi has minimum privacy violation issues and works under any lighting conditions. 
Besides, WiFi routers are prevalent in home/office spaces, thus the hardware infrastructure is already established in most indoor environments. 
In addition, WiFi-based ICC takes a cost-efficient device-free approach and does not require the targets to wear additional sensors. 
Due to the above-mentioned technical advantages, WiFi-based ICC is expected to be widely used in future wireless sensing applications. 

The existing WiFi-based ICC methods can be mainly divided into two categories, depending on the need of manual feature extraction \cite{zou2017freecount,liu2017wicount,liu2019deepcount,xi2020human,wang2021crowd,di2016trained,cheng2017device,zong2020phasedifference}. 
One relies on explicit manual features engineered from raw data, such as mean value, variance, median, and range, and then uses threshold-based or learning-based classifiers like support vector machine (SVM) to identify the crowd number. 
The other takes a fully data-driven approach and relies on deep learning models to extract the implicit features from the raw data measurements and performs crowd counting accordingly.  
The performance of the former method is critically related to the data feature selection.   
For example, to select the best-performing features, Zou et al. \cite{zou2017freecount} proposed a ``Transfer Kernel Learning (TKL)" method that selects data features based on a mutual information criterion from a feature pool including several statistical, transformation-based, and shape-based features. 
The performance of the latter method highly depends on iteratively training with a large number of labeled samples. For instance, the number of training samples  used by WiCount \cite{liu2017wicount} is more than 20000, which is difficult to implement in realistic application scenarios.

Therefore, despite the respective contributions of the above studies, the proposed methods suffer from a common drawback in practice. 
That is, although the well-trained deep learning models may achieve highly accurate ICC in one specific domain (even close to 100\% accuracy), once used in a new and dissimilar environment, e.g., different transceiver or background deployment, the cross-domain detection accuracy often plummets. 
For instance, our experiments show that the accuracy decreases sharply from 99\% to 12\% after applying a deep learning model trained in a rich-scattering office environment to a more spacious conference room.  
To achieve high ICC accuracy in a new domain, the above methods often require training their models from scratch. 
In practice, this is indeed infeasible because of the prohibitively high cost of collecting and labeling a large number of data samples for each new domain encountered. 
To facilitate large-scale deployment in the future, the WiFi-based ICC method must be able to achieve high classification accuracy across different domains even if only a very limited number of samples are available in cross-domain scenarios. 

In this article, we leverage the wisdom of few-shot learning (FSL) \cite{FewshotSurvey} to address the problems of model robustness and insufficient sample size in cross-domain ICC applications. 
In particular, we consider a practical scenario that the source domain has sufficient labeled training samples collected offline while the target domain only contains very few labeled samples. 
In this case, we propose a \textbf{D}omain-\textbf{A}gnostic and \textbf{S}ample-\textbf{E}fficient wireless crowd \textbf{C}ounting (DASECount) framework that can achieve high ICC accuracy in both source and target domains. 
To the authors’ best knowledge, this is the first work that leverages FSL to achieve robust ICC performance across different domains.
The main contributions of this paper are summarized as follows:  
\begin{itemize}
	
	\item We propose a DASECount framework for performing robust cross-domain ICC tasks. The DASECount framework includes two major stages: source domain meta training and target domain meta testing. 
	In the meta-training stage, a priori deep learning CNN model is trained on datasets collected in a local source domain to extract features from CSI amplitude and phase input data. 
	In the meta-testing stage, the well-trained CNN extracts the features of target domain data as the input to a tailor-made classifier, which eventually reports the final ICC result. 
	The DASECount framework is particularly useful as it requires as few as only 5 labeled data samples in a dissimilar target domain to reach over 99\% ICC accuracy.  
	\item In the source domain meta-training stage, DASECount devises two separate data pre-processing procedures for CSI amplitude and phase data, respectively. After preprocessing, it uses two CNN-based feature extractors to derive the low-dimension amplitude and phase features contained in the high-dimension input data, which facilitates training of the target domain classifier with very limited data samples. DASECount also applies a knowledge distillation technique to iteratively update the parameters of the CNN-based feature extractor, which improves at least 5\% ICC accuracy by experiments. 
	\item For target domain meta-testing, we first use the source domain feature extractor model to process the target domain training data. The output low-dimension features are considered as the input to train a lightweight classifier, e.g., a logistic regression model. By doing so, we can achieve high-performance cross-domain ICC even with very limited target domain training data size.
	\item We have conducted extensive experiments to evaluate the performance of the proposed DASECount framework in cross-domain ICC tasks. Results show that, with only $5$ labeled target domain training data samples per class, DASECount achieves accuracy of over 97\% in a 9-class ICC task when the crowd is stationary, over 99\% when the crowd moving randomly, and over 92\% in a more complex scenario with a mixture of stationary and moving crowds. We have also discussed the impact of detailed module designs in the proposed DASECount framework, e.g., selection of feature dimensions and classifier structures, on the cross-domain ICC performance. Overall, the proposed DASECount method attains robust and high accuracy in various cross-domain application scenarios.   
	
\end{itemize}

\begin{figure*}[!t]
	\centering
	\includegraphics[width=0.95\textwidth]{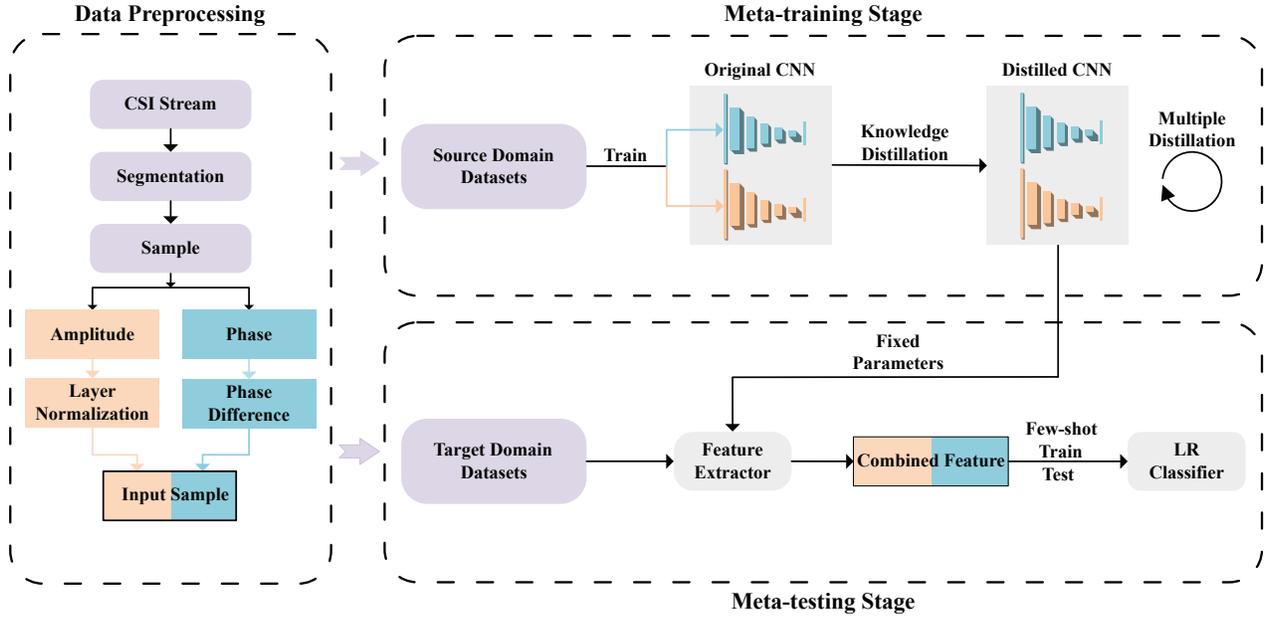}
	\caption{Schematics of the proposed DASECount cross-domain ICC method. The amplitude and phase of CSI measurements in both source and target domains are first preprocessed. In the meta-training stage, a CSI feature extractor, consisting of two CNN submodels (for processing amplitude and phase input), is trained on the source domain dataset , followed by a distillation process to refine the CNN model parameters. The parameters of the CNN submodels are fixed after source-domain training. In the meta-testing stage, the few labeled data samples in the target domain are first processed by the feature extractor. The combined amplitude-phase feature output is then used as the input for training a lightweight logistic regression (LR) classifier in a supervised manner.}
	\label{fig_system_architecture}
\end{figure*}

\section{Related Works}

\subsection{Learning-based WiFi ICC methods}
In recent years, deep learning methods have been widely used in IoT applications, such as smart cities \cite{8704334}, Internet of Vehicles \cite{9523794}, etc. 
For deep learning-based WiFi ICC methods, Liu et al. proposed WiCount \cite{liu2017wicount} model that implements a deep neural network (DNN)  for CSI-based crowd counting and achieves 82.3\% accuracy. The authors further improve the accuracy to 88.66\% with a new DeepCount model \cite{liu2019deepcount} that combines conventional neural network (CNN) \cite{CNN} and long short-term memory (LSTM) \cite{LSTM} structures.  
Xi et al. \cite{xi2020human} exploited CSI phase information and built a Resnet-based \cite{resnet} model, achieving on average 99\% accuracy of human counting. 
Wand et al. \cite{wang2021crowd} compared the performance of different deep learning networks, including CNN, LSTM, and gated recurrent unit \cite{GRU}, and showed that CNN achieves the best crowd counting accuracy.
 
Although human activity causes fluctuation of both amplitude and phase, many studies only use the amplitude information for ICC (like in \cite{di2016trained, cheng2017device, zou2017freecount}), mainly because the phase information often suffers from more severe hardware measurement noise such as carrier frequency offset and sampling time offset \cite{gong2016adaptive,palipana2016channel,zhu2017r}. 
Instead of using raw phase measurements, Zong et al. \cite{zong2020phasedifference} computed the phase difference between adjacent antennas as the input to an SVM-based ICC classifier. 
Liu et al. \cite{liu2020harvesting} utilized a CNN to extract the features of the CSI amplitude and phase information and use both features to detect human presence. It focuses on a binary classification problem which is a simplified special case of the general crowd counting problem considered in this paper.

\subsection{Few-shot Learning}
FSL and cross-domain algorithms were originally developed and applied in the field of computer vision and have now been extended to multiple application fields \cite{FewshotSurvey,MAML,MTL,rethinking,9350663,8826442,8906027}. There are several popular models to perform FSL. For instance, matching networks \cite{vinyals2016matching} encodes the data of the source domain and target domain into a feature space by learning an embedding function. Then, it compares the similarity between the two through cosine similarity to determine which category of the target domain data belongs to.
In recent years, researchers realized the importance of a priori models, that is, how a model that is fully trained and performs well on a task could be fine-tuned to handle a new task through learning with a limited number of samples.
\cite{MAML} proposed a meta-learning algorithm named ``MAML", which is suitable for many popular learning models that apply gradient descent for parameter training. 
The model trained by MAML can be efficiently fine-tuned with target domain data samples and it shows high classification accuracy even under a very limited target domain training data set. 
 
Different from fine-tuning global model parameters like MAML, \cite{MTL} proposed a new method called ``MTL'', which first trains a deep neural network (DNN) in the source domain. In the target domain, it fixes the general mass of neurons and fine-tunes the other neurons by a few-shot training sample of the target domain classification task. 
In a more recent work \cite{rethinking}, the authors took another FSL approach rather than fine-tuning the parameters of the well-trained model in the source domain. 
Instead, it utilized a pre-trained model as a feature extractor to process the target domain few-shot training samples for training a lightweight machine learning model.

\subsection{Applications of FSL to Wireless Sensing}
FSL methods have been practiced for CSI-based wireless sensing applications such as human activity recognition and gesture recognition. 
Shi et al. \cite{shi2020environment} proposed a MaNet-eCSI architecture using a matching network for CSI-based human activity recognition which can achieve a cross-domain recognition accuracy of 92.3\% with 5 training samples of the target domain. 
Zhang et al. \cite{zhang2020human} used MAML to train and fine-tune a 4-layer CNN for cross-domain human body activity recognition which reaches 89.6\% accuracy with 5 samples of new activity datasets.
\cite{zhang2021csi} proposed a human activity recognition model named ``CSI-GDAM'', which uses a convolutional block attention module \cite{woo2018cbam} layer to extract activity-related feature in CSI. CSI-GDAM reaches 99.74\% accuracy in 5-shot cases for cross-domain activity recognition. 
For CSI-based gesture recognition, Yang et al. \cite{yang2019learning} proposed a novel deep Siamese neural networks \cite{koch2015siamese} with multiple
kernel variant of maximum mean discrepancies \cite{gretton2012optimal} for cross-domain gesture recognition. The method can achieve an accuracy of 89.5\%  with only 1 sample in the target domain. 

It is worth noting that the above cross-domain wireless sensing methods mostly focus on fine-grained applications that classify human activities from a given set of patterns, such as a set of known gestures and body motions. 
In this case, the induced CSI variations are of a similar pattern and less sensitive to the background environment, thus high classification performance is likely achievable with a small set of training samples in the new domain.
In contrast to fine-grained applications, accurate cross-domain ICC tasks are more difficult because the CSI amplitude and phase variations caused by human free activities do not exhibit fixed patterns, instead are much more random and dependent on the domain environment. In this case, an ICC classifier trained with source domain data may face a severe over-fitting problem when applied to a new target domain. In this paper, we fully consider the unique challenge of cross-domain ICC tasks and propose a DASECount framework that provides robust cross-domain ICC performance.

\section{CSI Signal Model and Preprocessing Method}
In this section, we first introduce the WiFi sensing signal model and data format. Then we describe the CSI data pre-processing method to prepare input data for the cross-domain ICC tasks of the DASECount framework as shown in Fig. \ref{fig_system_architecture}.

\subsection{CSI Signal Model} \label{introduction}
In WiFi communication, channel state information (CSI) reflects the signal variations during transmission between the transmitter and receiver, including channel amplitude attenuation and phase shift \cite{r01}. The channel frequency response described by CSI is
\begin{equation}
	\widetilde{H}(f;t)=\sum_{n=1}^{N}a_n(t)e^{-j2\pi f\tau_n(t)},
\end{equation}
where $N$ represents the number of multipaths,  $a_n(t)$ and $\tau_n(t)$ represent the amplitude attenuation and propagation delay in the $n$th path, and $f$ denotes the carrier frequency. The receiving signal could be described as
\begin{equation}
	Y(f;t)=\widetilde{H}\cdot X(f;t) + n(f;t),
\end{equation}
where $X(f;t)$ is the transmitting signal in frequency $f$ and at time $t$, $Y$ is the conrresponding received signal, and $n$ is the receiver noise. 
 
IEEE 802.11a/b/n WiFi protocol supports multiple-input multiple-output (MIMO) and orthogonal frequency-division multiplexing (OFDM) transmissions.
CSI acquisition requires specialized software operating on particular WiFi card chips.  
Two popular WiFi CSI acquisition softwares are the Intel 5300 CSI Tool \cite{Intel5300} and the Atheros CSI Tool \cite{AtherosCSI}. 
While the former supports 20Mhz bandwidth 30 subcarriers, the latter supports two operating modes: 20Mhz bandwidth 56 subcarriers and 40Mhz bandwidth 114 subcarriers. 
In this paper, we use the Atheros CSI tool to collect the CSI between a pair of transceivers with 2 transmitting and 3 receiving antennas, operating at 40MHz with 114 subcarriers. 
In this case, the collected CSI data is expressed as a 4-dimensional complex tensor $\bar{H}\in\mathbb{C}^{T\times N_r \times N_t \times N_{sc}}$, where $T$, $N_r$, $N_t$, $N_{sc}$ are the number of time frames, receiving antennas, transmitting antennas, and subcarriers, respectively. 
%Similar to image classification methods, CSI could utilize deep learning method such as CNN to tackle with ICC classification problem.

\subsection{Proposed CSI Preprocessing Method} \label{preprocess}
\begin{figure}[H]
	\centering
	\includegraphics[width=0.5\textwidth]{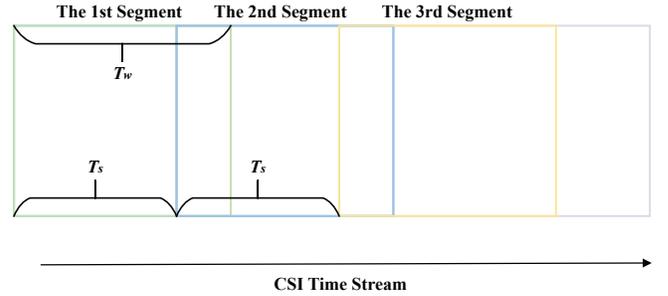}
	\caption{Illustration of the data segmentation process. The shaded parts in the figure represent the overlap between two segments.}
	\label{fig_segment}
\end{figure}

To facilitate subsequent processing by machine learning models, we propose the following preprocessing procedures on the collected CSI tensor data. As shown in Fig. \ref{fig_segment}, we first use a slide window of duration $T_s$ to split the raw data into equal segment of duration $T_w$. The resulting CSI data within a tagged segment is expressed as  $\widetilde{H}\in\mathbb{C}^{T_w\times N_r \times N_t \times N_{sc}}$. 
Here, we set $T_s < T_w$ to produce an overlap $T_w-T_s$ between the two adjacent segments, which brings two benefits: increase the number of training samples after segmentation and traverse the CSI variations caused by human movements in different periods. 
Then, we extract the amplitude data $\widetilde{H}^{amp}$ and phase data $\widetilde{H}^{pha}$ from each complex CSI data segment $\widetilde{H}$.

It is a common practice to impose noise reduction methods, such as Hampel \cite{zhu2017r} and low-pass filters \cite{oshiga2019human}, to amplitude data for fine-grained application scenarios like gesture recognition and respiratory monitoring.
However, for a coarse-grained application like ICC, our empirical results show that amplitude noise reduction may lead to severe performance degradation as it may falsely remove the random high-frequency signal variations caused by simultaneous movements of multiple people.  
Therefore, we use the raw CSI amplitude data without applying the noise reduction technique. 

%\emph{CSI amplitude:}
Here, we first rearrange the amplitude data to a dimension of $N_{rt} \times T_w\times N_{sc}$ , where $N_{rt} = N_r \cdot N_t$ denotes the number of parallel CSI between the transmit and receive antennas.
Then, we process each $\widetilde{H}^{amp}$ with Layer Normalization \cite{ba2016layer}. 
Specifically, we denote $\hat{a}_{l,i,j}$ as the amplitude measurement corresponds to the $l$th Tx-Rx antenna pair, the $i$th time slot and the $j$th sub-carrier. 
Then, we compute the mean $\mu_l$ and the standard deviation $\sigma_l$ of the $T_w \times N_{sc}$ amplitude measurements taken from the $l$th antenna pair. 
The normalization method for each amplitude measurement is expressed as 
\begin{equation}
	a_{l,i,j}=\frac{\hat{a}_{l,i,j}-\mu _l}{\sigma _l}, \forall l,i,j.
\end{equation}
After layer normalization, we denote the amplitude data as $H^{amp}$.  

%\emph{CSI phase:}
For phase data processing, due to hardware impairment of WiFi chips, such as carrier frequency offset, and sampling time offset \cite{liu2020harvesting}, the CSI phase data often change abruptly in adjacent time slots. 
Here, we first use the ``unwrap" function to correct phase jump, and then compute the phase difference between two adjacent receiving antennas to eliminate random phase noise \cite{zong2020phasedifference}, and denote the phase data after processing as  $H^{phd}$.

With a bit abuse of notation, we denote the data samples in the $i$th segment as  denoted as $x_i=(H_i^{amp}, H_i^{phd})$, where $H_i^{amp}$ is the CSI amplitude part and $H_i^{phd}$ is the CSI phase difference part. 
For each measurement $x_i$, we append a label $y_i \in\{0,1,\cdots, M\}$ denoting the number of people in the test, where $M$ denotes the maximum number of people considered. 

\begin{figure*}[!t]
	\centering
	\includegraphics[width=0.9\textwidth]{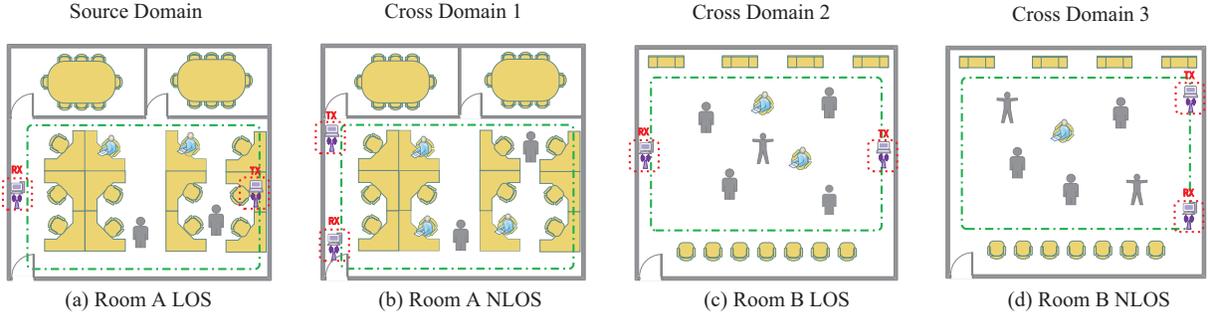}
	\caption{Data collection scenarios. Data collection is conducted in 2 rooms, each containing LOS and NLOS scenarios.}
	\label{fig_scene}
\end{figure*}

\begin{table*}[htbp]
	\centering
	\caption{Nomenclature}
	\begin{tabular}{ c | c | c }
		\toprule
		Symbol                         & Terminology                            & Description  \\
		\midrule
		$\mathcal{S}$                  & Source domain dataset                  & Contains multiple ICC tasks of the source domain    \\
%		\midrule
%		$s$                            & Source domain ICC task                 & Motion types of crowd in the source domain scenarios     \\
		\midrule
		$\mathcal{T}$                  & Target domain dataset                  & Contains multiple ICC tasks of the target domain   \\
%		\midrule
%		$t$                            & Target domain ICC task                 & Motion types of crowd in the target domain scenarios     \\
		\midrule
		$\mathcal{D}^{train}$        & Training set of $\mathcal{S}$            & Training samples of an ICC task in the source domain  \\
		\midrule
		$\mathcal{D}^{val}$          & Validating set of $\mathcal{S}$          & Validating samples of an ICC task in the source domain  \\
		\midrule 
		$\mathcal{D}^{sup}$          & Support set of $\mathcal{T}$             & Few-shot samples of an ICC task of the target domain, used to train a classifier   \\
		\midrule
		$\mathcal{D}^{que}$          & Query set of $\mathcal{T}$               & Used for evaluating the performance of the target domain classifier  \\
		\midrule
		$x_*$                          & Input sample                           & Includes the CSI amplitude part ${H_*^{amp}}$ and the CSI phase difference part $H_*^{phd}$ \\
		\midrule
		$y_*$                          & Label                                  & The ground-truth number of people in the scene   \\
		\midrule
		$\phi$                          & CNN parameters                        & Contains amplitude and phase difference submodel $\phi^{amp}$, $\phi^{phd}$  \\
		\midrule
		$\psi$                          & Feature extractor                      & Generated by using partial parameters of $\phi$   \\
		\midrule
		$\theta$                          & Classifier                     & For specific ICC tasks in the target domain   \\	
		\bottomrule
	\end{tabular}%
	\label{tab1}%
\end{table*}%

\section{The Proposed DASECount Framework}
In this section, we introduce the DASECount framework for cross-domain ICC tasks. 
We divide the CSI data into source domain datasets and target domain datasets, respectively. 
The source domain dataset contains a large number of labeled sample sets collected from a local pre-set scene, while the target domain data set contains limited labeled data samples collected from the scene to be detected. 
For ICC tasks, CSI is often sensitive to the deployment of the WiFi transceivers and the surrounding environment. 
Without loss of generality, we consider a particular equipment deployment and room environment as the source domain scenario, and any significant change of equipment deployment or environment from the target domain leads to a new target domain scenario.  

An example source-target domain setup is illustrated in Fig. \ref{fig_scene}. 
We consider two rooms where Room A is a rich scattering office room and Room B is a spacious conference venue. 
Besides, we also consider both line-of-sight (LOS) and non-line-of-sight (NLOS) WiFi equipment placements, where the human targets are in the LOS and NLOS channels of the WIFi transceivers, respectively.
In total, there are four different scenarios, we consider without loss of generality that Room A LOS case as the source domain, and the rest three as target domains.

As shown in Fig. \ref{fig_system_architecture}, after data collection and preprocessing, the proposed DASECount framework contains two main stages: the meta-training stage and the meta-testing stage. We will describe each stage below. The symbols involved are shown in Table \ref{tab1}.
%In the meta-training stage, a CSI feature extractor, consisting of two CNN submodels (for processing amplitude and phase difference data), is trained on the source domain dataset for extracting the implicit features of target domain data. The parameters of the CNN submodels are fixed after source-domain training. 
%
%In the meta-testing stage, data samples in the target domain are first processed by the feature extractor to produce amplitude and phase data features. 
%The combined amplitude-phase features are then used for training a lightweight classifier (e.g., an LR-based classifier) on the target domain samples. 
%In the following subsections, we introduce in details the meta-training and meta-testing stages.   
\subsection{Source Domain Meta-training Stage} \label{meta-training}
We denote the source domain dataset as $\mathcal{S}={(\mathcal{D}^{train},\mathcal{D}^{val})}$, where $\mathcal{D}^{train}=\{\mathcal{D}_s^{train}\}_{s=1}^S \triangleq  (x_i,y_i)_{i=1}^I$ is the training set, $\mathcal{D}^{val}=\{\mathcal{D}_s^{val}\}_{s=1}^S \triangleq  (\hat{x}_j,\hat{y}_j)_{j=1}^J$ is the validating set. Here, $s$ represents the $s$th type of ICC task in the source domain. For example, in the simulation section, we consider $S=3$ types of ICC tasks, where $s=\{1,2,3\}$ corresponds to ICC tasks when the targets are under static, dynamic, and a mixed static and dynamic motions, respectively. Besides, $I$ and $J$ denote the total number of data samples used for training and validation, respectively. The detailed descriptions are presented in Section \ref{device}.

In the meta-training stage, we first train a CSI feature extractor using the source domain dataset $\mathcal{S}$ by supervised learning. We denote the model parameters of the feature extractor as $\phi$, which contains two sub-models, one processes CSI amplitude information denoted as $\phi^{amp}$ and the other processes CSI phase difference information denoted as $\phi^{phd}$. Different from \cite{liu2020harvesting}, where amplitude and phase modules are concatenated by a fully connected layer, the amplitude and phase difference submodels of DASECount are independent and described below. 
%In order to improve the robustness of the feature extractor and reduce training complexity, we build a merged training set $\mathcal{D}^{train}=\{\mathcal{D}_s^{train}\}_{s=1}^S$ and a merged validating set $\mathcal{D}^{val}=\{\mathcal{D}_s^{val}\}_{s=1}^S$ to train and evaluate a unified feature extractor instead of training a separate one for each ICC task. 

\begin{figure}[!t]
	\centering
	\includegraphics[width=0.5\textwidth]{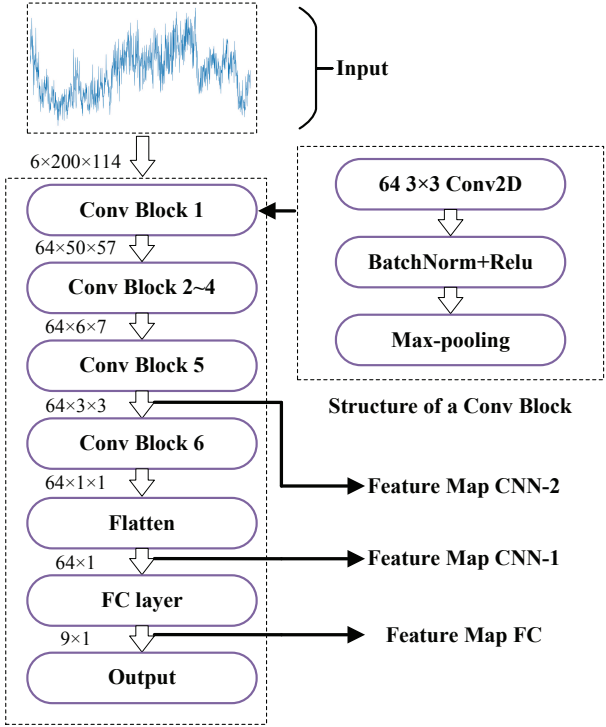}
	\caption{The structure of the CNN stacks. The amplitude and phase difference submodels have the same CNN structure in the figure. }
	\label{fig_cnn}
\end{figure}

To fully exploit both the chronological and subcarriers correlations, we apply a 2D CNN consisting of 6 convolutional blocks and a fully connected layer, as shown in Fig. \ref{fig_cnn}. 
Each convolutional block contains a convolutional layer (64 $3\times 3$ convolution 2D kernels), a batch normalization layer, a Relu activation function, and a max-pooling layer. 
The number of neurons in the fully connected layer is the same as the number of sample classes. 
It is worth mentioning that the first pooling layer has a kernel of $4\times 2$ and all the others have a kernel of $2\times 2$. 

The process of training $\phi$ is expressed as follows:
\begin{equation} 
	\phi=\mathop{\arg}\min_{\phi}\mathcal{L}^{ce}(\mathcal{D}^{train};\phi),
\end{equation}
where $\mathcal{L}^{ce}$ represents the cross-entropy loss between source domain data and corresponding labels. We train the $\phi$ on the training set $\mathcal{D}^{train}$ and evaluate it on the validating set $\mathcal{D}^{val}$. We present the training procedures of the amplitude and phase submodels in Algorithm \ref{alg_feature extractor}. 
%Specifically, we combine all the motion types samples to build a incorporative set and randomly divide it into $\mathcal{D}_{train}^{new}$ and $\mathcal{D}^{val}$ in a ratio, which is detailed in \ref{train cnn}. 
%Hence, $\mathcal{D}_{train}^{new}$ contains all the training dataset in the source domain, i.e., 
%\begin{equation} \label{eq_D_new}
%	\mathcal{D}_{train}^{new}=\{\mathcal{D}_s^{train}\}_{s=1}^S  \triangleq  (x_i,y_i)_{i=1}^I.
%\end{equation}
%while $\mathcal{D}^{val}$ contains all the validating dataset in the source domain, i.e.,
%\begin{equation} \label{eq_D_new_val}
%	\mathcal{D}^{val}=\{\mathcal{D}_s^{val}\}_{s=1}^S  \triangleq  (\hat{x}^j,\hat{y}^j)_{j=1}^J.
%\end{equation}

\begin{algorithm}[htb] 
	\caption{ Training Procedures of the Feature Extractor } 
	\label{alg_feature extractor} 
	\begin{algorithmic}[1]
		\REQUIRE ~~\\ % Input
		Merged training set $\mathcal{D}^{train}$;
		\ENSURE ~~\\ % Output
		Feature extractor $\phi = (\phi^{amp}, \phi^{phd}) $;\\
		
		%\STATE // Train feature extractor  $\phi^{amp}, \phi^{phd} $ on $\mathcal{D}^{train}$
		%\WHILE{not none}
		
		\STATE  Initialize model parameters $\phi^{amp}, \phi^{phd}$, learning rate $\eta$.
		
		\FORALL{$H_i^{amp}$ of $ x_i \in \mathcal{D}^{train}$}
		\STATE  Calculate the output $\hat{y}_i^{amp} = f(H_i^{amp};\phi^{amp})$ 
		\STATE  Calculate $\mathcal{L}^{ce}(\hat{y}_i^{amp} , y_i)$ \
		\STATE  Update $\phi^{amp}=\phi^{amp}-\eta\nabla_{\phi^{amp}}\mathcal{L}^{ce}(\hat{y}_i^{amp} , y_i)$ \
		\ENDFOR
		
		\FORALL{$H_i^{phd}$ part of $ x_i \in \mathcal{D}^{train}$}
        \STATE  Calculate the output $\hat{y}_i^{phd} = f(H_i^{phd};\phi^{phd})$ 
		\STATE  Calculate $\mathcal{L}^{ce}(\hat{y}_i^{phd} , y_i)$ \
		\STATE  Update $\phi^{phd}=\phi^{phd}-\eta\nabla_{\phi^{phd}}\mathcal{L}^{ce}(\hat{y}_i^{phd} , y_i)$ \
		\ENDFOR
		\STATE  Output $\phi^{amp}, \phi^{phd}$

	\end{algorithmic}
\end{algorithm}

To improve the generalization capability of the feature extractor, we apply the knowledge distillation technique \cite{hinton2015distilling}, which has shown effective performance improvement of FSL problem for image classification \cite{rethinking}. 
We treat the original trained CNNs in Algorithm \ref{alg_feature extractor} as the initial teacher model $\phi_0$, and iteratively distill knowledge from the teacher model to a student model using the same source domain training dataset. 

\begin{figure}[H]
	\centering
	\includegraphics[width=0.5\textwidth]{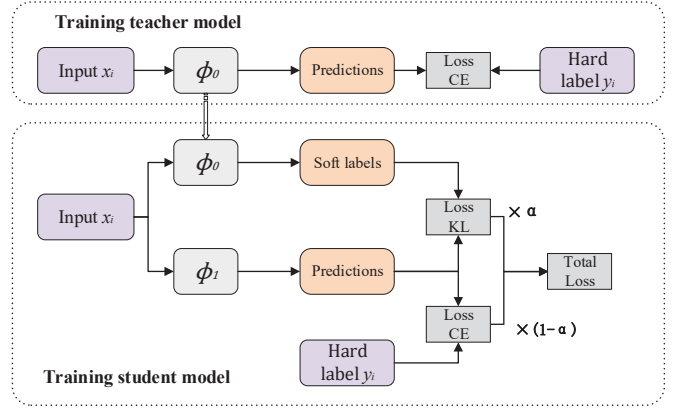}
	\caption{The model knowledge distillation procedure. The figure shows the process of training the $1$st generation distilled model. The total loss is used to update the model parameters of $\phi_1$.}
	\label{fig_distillation}
\end{figure}

In the proposed DASECount framework, the teacher and student models have the same structure, and we use a fixed training set $\mathcal{D}^{train}$ as the input to all the models.
As shown in Fig. \ref{fig_distillation}, suppose that the input for training models is $x_i$ and the corresponding label is $y_i$. 
Let $f(x_i;\phi_0)$ denote the output of the initial teacher network given an input $x_i$.
To obtain the first distilled model $\phi_1$, we use $y_i$ as the hard label and $f(x_i;\phi_0) $ as the soft label of the input $x_i$. 
Similarly, to generate the $k$th distilled model $\phi_{k}$, we solicit both the hard label $y_i$ and the soft label $f(x_i;\phi_{k-1})$ to the input $x_i$ and minimize the weighted loss caused by both the hard and soft labels \cite{furlanello2018born}.
Therefore, the distillation of the $k$th model can be written as follows:
\begin{equation} \label{eq_distillation}
	\begin{split}
		\phi_k = \arg\mathop{\min}_{\phi} &(\alpha\mathcal{L}^{ce}(\mathcal{D}^{train};\phi)+\\
	&(1-\alpha) KL(f(\mathcal{D}^{train};\phi),f(\mathcal{D}^{train};\phi_{k-1}))), \\
	\end{split}
\end{equation}
where $\alpha \in [0,1]$ is the weight of cross-entropy (CE) loss, $KL$ represents the Kullback-Leibler divergence between two distributions. 
Both the amplitude and phase CNN models are distilled several times. 
The parameter update of model distillation is shown in Algorithm \ref{alg_distillation}.
After distillation, we have obtained a series of distilled model $ \{\phi_k\}_{k=0}^{K} $. 
We choose one model $\phi_m = (\phi_m^{amp}, \phi_m^{phd})$ to generate the final CSI feature extractor, denoted as $g_\psi$, where $\psi$ represents the model parameters.
We will demonstrate the selection method of the distilled models and the advantage of distillation to the cross-domain ICC performance in Section \ref{effect of distillation}. 

\begin{algorithm}[htb]
	\caption{ Distillation Procedure of Feature Extractor } 
	\label{alg_distillation} 
	\begin{algorithmic}[1]
		\REQUIRE ~~\\ % Input
		Merged training set $\mathcal{D}^{train}$; \\
		Initial feature extractor $\phi_0 = (\phi_0^{amp}, \phi_0^{phd})$;
		\ENSURE ~~\\ % Output
		
		$(\phi_k^{amp}, \phi_k^{phd})_{k=0}^K $;\\
		
		%\STATE  Initialize $\phi_{(0)}=\phi$
		
		\FOR {$k$=1:$K$}
		\STATE  Initialize the $k$th model ${\phi_k} = (\phi_k^{amp}, \phi_k^{phd})$
		
		\FORALL {$H_i^{amp}$ of $ x_i \in \mathcal{D}^{train}$}
		\STATE Calculate the output $f(H_i^{amp};\phi_{k}^{amp})$
		\STATE Obtain soft label $f(H_i^{amp};\phi_{k-1}^{amp})$
		%		\STATE Calculate $KL(\phi_k^{amp}(H_i^{amp}), \phi_{k-1}^{amp}(H_i^{amp}))$
		%		\STATE Calculate $\mathcal{L}^{ce}(\phi_k^{amp}(H_i^{amp}), y_i)$
		%		\STATE Calculate total loss $\mathcal{L}^{all} = \alpha\mathcal{L}^{ce} + (1-\alpha)KL$
		\STATE Update $\phi_k^{amp} $ by the Equation (\ref{eq_distillation})
		\ENDFOR
		
		\FORALL {$H_i^{phd}$ of $ x_i \in \mathcal{D}^{train}$}
		\STATE Calculate the output $f(H_i^{phd};\phi_{k}^{phd})$
		\STATE Obtain soft label $f(H_i^{phd};\phi_{k-1}^{phd})$
		%		\STATE Calculate $KL(\phi_k^{phd}(H_i^{phd}), \phi_{k-1}^{phd}(H_i^{phd}))$
		%		\STATE Calculate $\mathcal{L}^{ce}(\phi_k^{phd}(H_i^{phd}), y_i)$
		%		\STATE Calculate total loss $\mathcal{L}^{all} = \alpha\mathcal{L}^{ce} + (1-\alpha)KL$
		\STATE Update $\phi_k^{phd} $ by the Equation (\ref{eq_distillation})
		\ENDFOR
		
		\ENDFOR

	\end{algorithmic}
\end{algorithm}

\subsection{Target Domain Meta-testing Stage} \label{meta-testing}
After obtaining the distilled CSI feature extractor in the source domain, we select the output of a particular layer as the feature to train the target domain classifier. For instance, as illustrated in Fig. \ref{fig_cnn}, the choice can be the feature map of FC, CNN-1, CNN-2, etc, and we leave the discussion of feature map selection in Section \ref{FE_Select}. 
For the target domain, we denote the dataset as $\mathcal{T}={(\mathcal{D}^{sup},\mathcal{D}^{que})}_{t = 1}^T$, 
where $\mathcal{D}^{sup} = \{\mathcal{D}_t^{sup}\}_{t=1}^{T} \triangleq (x_t^p, y_t^p)_{p=1}^P $ is the support set,  $\mathcal{D}^{que} = \{\mathcal{D}_t^{que}\}_{t = 1}^T  \triangleq (x_t^q, y_t^q)_{q=1}^Q$ is the query set and $t$ represents the $t$th ICC task in the target domain. 
Because the support set contains very limited labeled training samples, we use a shallow classifier parameterized by $\theta$. 
Some example classifiers include logistic regression (LR) and support vector machine (SVM), etc. 
In particular, we minimize the cross-entropy loss with the FSL training samples of the support set:
\begin{equation}
    \theta=\mathop{\arg}\min_{\theta}\mathcal{L}^{ce}(\mathcal{D}^{sup};\theta).
\end{equation}
Using LR as the classifier, the meta-testing procedure in the target domain is shown in Algorithm \ref{alg_classifier}. 
Finally, we evaluate the performance of the classifier in the query set $\mathcal{D}^{que}$.

\begin{algorithm}[htb] 
	\caption{ Training procedure of target domain classifier } 
	\label{alg_classifier} 
	\begin{algorithmic}[1]
		\REQUIRE ~~\\ % Input
		Support set $\mathcal{D}^{sup}$ of an ICC task in the target domain; \\
		Feature extractor $g_\psi$;
		\ENSURE ~~\\ % Output
		An LR classifier $\theta$ for the ICC task;\\
		
		\FORALL {few-shot samples $x_t^p \in \mathcal{D}_t^{sup}$}
		\STATE Compute feature map $\Phi_p=g_\psi(x_t^p)$ \
		\STATE Reshape feature map to 1D vector and augment 5 times \
		\STATE Training the LR classifier:  
		\STATE $\theta = \theta - \eta(-\frac{1}{P} \sum_{p=1}^{P} [(y_t^p - \frac{1}{1 + e^{-W^T \Phi_p}})\Phi_p])$ \
		\ENDFOR

	\end{algorithmic}
\end{algorithm}

\section{Experiment Results}
In this section, we perform experiments to evaluate the performance of the proposed DASECount framework. 
We first describe the experiment setups and parameter settings in Section \ref{device}. 
Then, we show the performance of the CNN-based feature extractor in handling source domain ICC tasks Section \ref{train cnn}. 
In Section \ref{result}, we present the results of cross-domain ICC tasks, where we compare DASECount with other benchmark methods and discuss design factors that influence the performance. 

\subsection{Experiment Setups} \label{device}
We use a laptop as the transmitter and a desktop as the receiver, where both devices communicate with Atheros 802.11n WiFi card (AR9580/AR9382). 
Meanwhile, both devices run Ubuntu 14.04 system and the receiver uses the Atheros CSI tool to collect the CSI. 
The transmission works in the 2.4GHz spectrum, occupying 40MHz bandwidth with 114 subcarriers, using 2 transmitting antennas and 3 receiving antennas. 
The transmit rate is set to 100 pkts/s. All the collected data packets are parsed by the Atheros CSI tools and further processed by MATLAB. All the data processing and computations are performed on a Dell PowerEdge T640 server with 256GB of RAM and a Tesla P100 GPU. 

For cross-domain ICC task setups, we conduct experiments in 2 rooms where each room contains a line-of-sight (LOS) and a none-line-of-sight (NLOS) experimental scenario, resulting 4 different scenarios in total. 
As shown in Fig. \ref{fig_scene}, room A is an office space while room B is a lecture hall. The considered cross-domain scenario setup is similar to that in \cite{gao2021ml}.

To measure the effects of different types of human activity on performance, we consider the following three motion types and collect data for each type when 0 to 8 (i.e., 9 classes) volunteers are in the room. 
\begin{itemize}
	\item Static: volunteers are required to remain seated but can act freely, such as eating, typing, or sleeping;
	\item Dynamic: volunteers walk randomly walk around the venue; 
	\item Mixed: there is no restriction to the volunteers' activities, and they can move freely in the venue including but not limited to walking, sitting, eating, and sleeping. 
\end{itemize}

CSI data of each category (4 scenarios $\times$ 3 motion types $\times$ 9 classes = 54 categories in total) are collected for 5 minutes to obtain a total of about 30000pkt for each category. 
In the CSI pre-processing stage, we set segmentation window $T_w$ as 200 (i.e., 2 seconds as a unit) and sliding window $T_s$ as 50 (i.e., 0.5 seconds), so we obtain 600 segments for each category.

After preprocessing, we have obtained CSI data of 3 motion types in each scenario, where each scenario-motion pair has 5400 samples (600 samples $\times$ 9 classes). 
We treat these different motion types as different ICC tasks in each scenario (i.e., subscript $s$ for source domain and $t$ for target domain in \ref{meta-training}). 
Without loss of generality, data samples collected under Room A LOS scenario are used as source domain dataset $\mathcal{S}$ and samples in other 3 scenarios are used as target domain datasets $\mathcal{T}$. 

\subsection{Feature Extractor Configuration} \label{train cnn}
We generate the feature extractor in the source domain dataset $\mathcal{S}$, which contains data of all the three motion types. 
We train a unified feature extractor rather than one for each of the 3 motion types. 
Therefore, we combine the samples of all motion types to build a united dataset, and then divide it into training set (i.e., $\mathcal{D}^{train}$ in \ref{meta-training})) and validating set (i.e., $\mathcal{D}^{val}$ in \ref{meta-training}) in a ratio of 9 to 1. 
Hence, the training set $\mathcal{D}^{train}$ contains $3 \times 9 \times 540 = 14580$ samples and the validating set $\mathcal{D}^{val}$ contains $3 \times 9 \times 60 = 1620$ samples. 
The training of the feature extractor is carried out on $\mathcal{D}^{train}$. 
Some main training parameters involved are shown in Table \ref{tab2}. Both the amplitude and phase difference submodels are trained with the same training parameters. 

\begin{table}[htbp]
	\centering
	\caption{Training parameters of feature extractor}
	\begin{tabular}{  c |c | c| c }
		\toprule
	              Batch size    &Epochs   & Learning rate      &Optimizer  \\
		\midrule
                      8         &30       & $10^{-3}$             & Adam \\
		
		\bottomrule
	\end{tabular}%
	\label{tab2}%
\end{table}%

\begin{figure}%[H]
	\centering  
	\subfloat[Amplitude Submodel]{
		\label{Fig.sub.12}
		\includegraphics[width=0.45\textwidth]{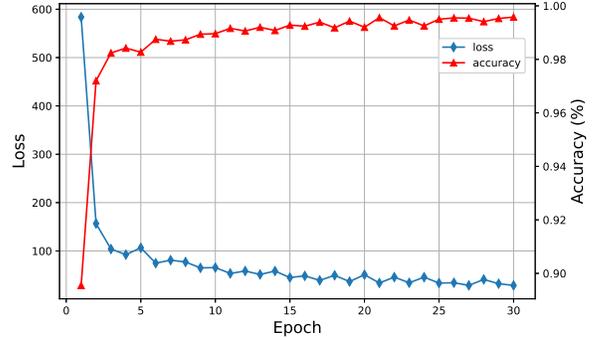}}
	\hfil
	\subfloat[Phase difference Submodel]{
		\label{Fig.sub.22}
		\includegraphics[width=0.45\textwidth]{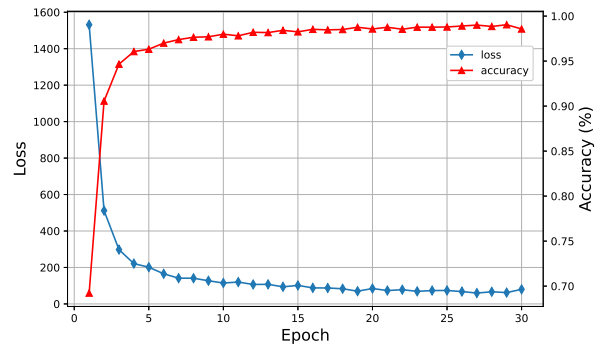}}
	\caption{Training loss and ICC accuracy of the amplitude and phase difference submodels in the meta-training stage.}
	\label{fig_loss}
\end{figure}

After training, we evaluate the ICC performance of the amplitude and phase difference submodels on the source domain validation set $\mathcal{D}^{val}$. 
In Fig. \ref{fig_loss}, we show the loss and accuracy variations in 30 training epochs. 
As can be seen from Fig. \ref{Fig.sub.12}, the training loss of the amplitude submodel drops rapidly in the first 10 epochs and gradually converges after 30 epochs. The accuracy of the model reaches 98\% on the validation set. 
The loss and accuracy of the phase difference submodel vary similarly to the amplitude submodel during the training process.

Besides, we perform knowledge self-distillation 6 times to the original feature extractor, obtaining 7 generation models in total. Some main distillation parameters involved are shown in the table \ref{tab3}. Cross entropy loss weight $\alpha$ in Equation (\ref{eq_distillation}) is set to 0.5. 
Among these models, we select a model with the best ICC accuracy as the target domain feature extractor, i.e., the 4th generation model.% as the target domain feature extractor.

\begin{table}[htbp]
	\centering
	\caption{Distillation parameters of feature extractor}
	\begin{tabular}{ c | c | c| c | c}
		\toprule
		     Batch size &Epochs & Learning rate   & Optimizer  & Weight decay \\
		\midrule
		       100 & 100 & $10^{-3}$    & SGD & $5\cdot 10^{-4}$  \\
		
		\bottomrule
	\end{tabular}%
	\label{tab3}%
\end{table}%

\subsection{Performance of DASECount} \label{result}
In the meta-testing stage, $k$-shot samples in the target domain are input to the obtained extractor to compute the corresponding feature maps, where $k \in \{1,5\}$ correspond to 1-shot and 5-shot cases, respectively. 
Here, we choose the feature maps from the penultimate convolutional block to train and evaluate the target domain LR classifier with 1-shot learning and 5-shot learning (the feature map CNN-2 in Fig.\ref{fig_cnn}). 
As shown in Fig. \ref{fig_cnn}, the $64\times 3\times 3$ feature map is flattened to a $1\times 576 $ vector. 
After concatenating features from the amplitude and the phase difference submodels, we have obtained a $1\times 1152$ feature vector. 
Then, each $k$-shot sample vector is duplicated five times. 
For 1-shot learning, only 1 training sample is available for each class of the 9 classes. 
After duplication, the dimension of training data is $45\times 1152$. 
For 5 shot learning, 5 samples are available for each class so that the dimension of training data is  $225\times 1152$, accordingly.

\subsubsection{ICC accuracy with LR classifier} \label{result1}
Table \ref{tab_result} shows 1-shot and 5-shot results of the LR classifier with the 4th generation distillation feature extractor. 
All the accuracy result is an average obtained by repeating the experiment 10 times.

\begin{table}%[H]
	\centering
	\caption{Few-shot results for meta-testing scenarios}
	\begin{tabular}{ c | c c c c }
		\toprule
		&                                   & Room A NLOS & Room B LOS & Room B NLOS \\
		\midrule
		\multirow{3}{*}{1 shot} & Static    & 90.05\%     & 88.63\%    & 85.64\% \\
	                         	& Dynamic   & 94.63\%     & 93.41\%    & 93.29\% \\
		                        & Mixed     & 80.65\%     & 77.17\%    & 76.26\% \\
		\midrule
		\multirow{3}{*}{5 shot} & Static    & 98.29\%     & 97.83\%    & 97.26\% \\
	                            & Dynamic   & 98.33\%     & 98.94\%    & 99.17\% \\
		                        & Mixed     & 96.85\%     & 94.61\%    & 92.68\% \\
		\bottomrule
	\end{tabular}%
	\label{tab_result}%
\end{table}%

We see that, compared with 1-shot learning, the accuracy improvement of 5-shot learning ranges from the lowest 4\% (Dynamic type of Room A NLOS) to the highest 16\% (Mixed type of Room B NLOS), which matches the intuition that increasing the number of shots improves detection performance.
For the Room A NLOS scenario, the LR classifier achieves an average accuracy of 97.82\% with 5-shot learning. The high accuracy is because only the transceiver deployment is changed compared to the source domain. 
As for the Room B NLOS scenario, the average detection accuracy drops slightly (about 1.5\%) because of the change of the entire surrounding environment. 
From the perspective of motion types, the accuracy is higher (average 98.81\% with 5-shot learning) with the Dynamic type and lower (average 94.71\% with 5-shot learning) with the Mixed type, because of the higher degree of randomness in the CSI measurements under the Mixed motion type of the crowd. 

%For 1 shot learning, the LR classifier obtains an accuracy of 76.26\% in Mixed motion mode of Room B NLOS, which is the most difficult scenario. 
%The results show that the LR classifier trained by few-shot samples, generally has a excellent good performance, even in the case of mixed motion modes, which is the most difficult to detect. 

\begin{figure*}
	\centering
	\subfloat[Confusion Matrix of Static type]{
		\label{Fig.sub.14}
		\includegraphics[width=0.30\textwidth]{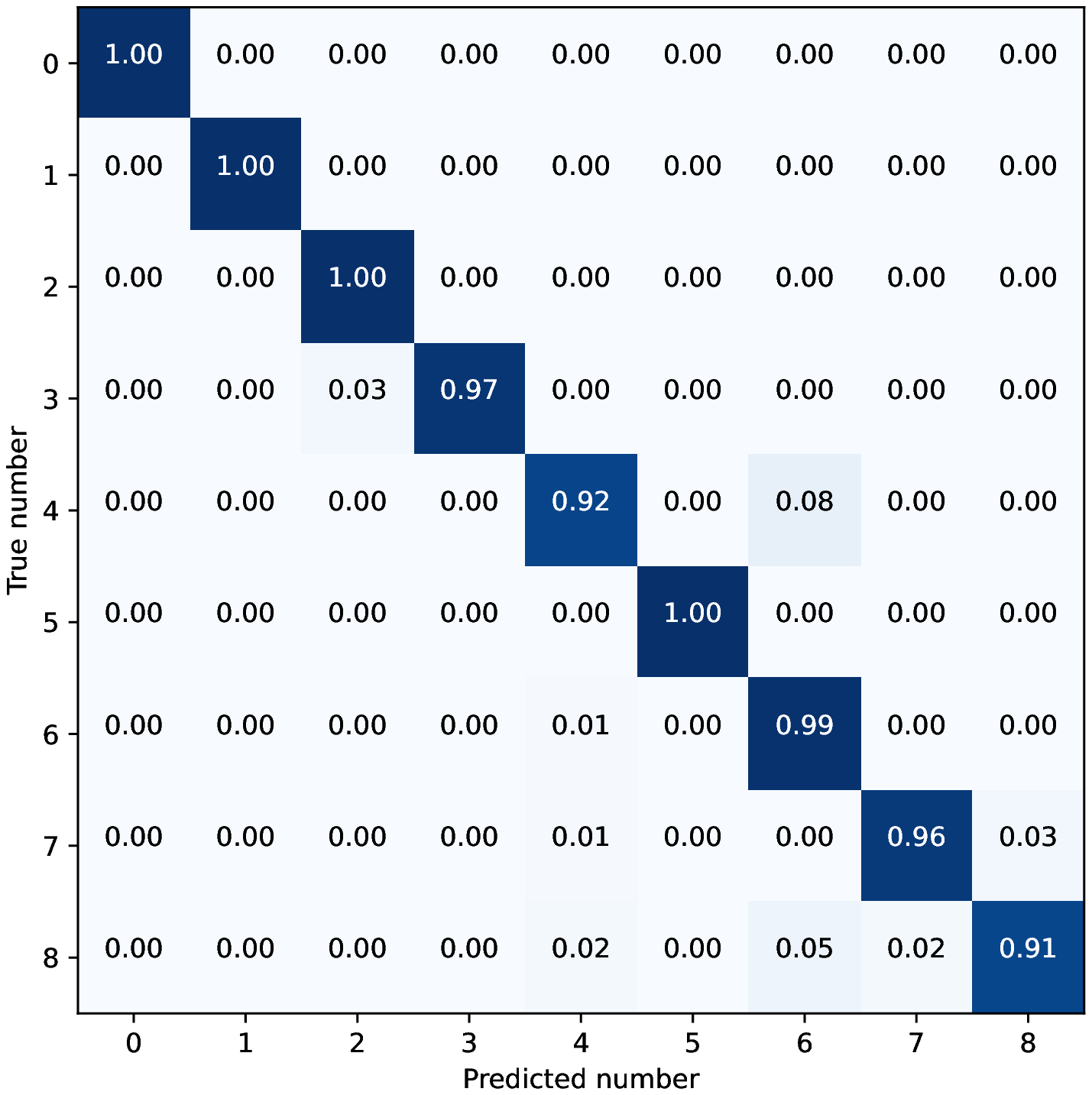}}
	\hfil
	\subfloat[Confusion Matrix of Dynamic type]{
	    \label{Fig.sub.24}
     	\includegraphics[width=0.30\textwidth]{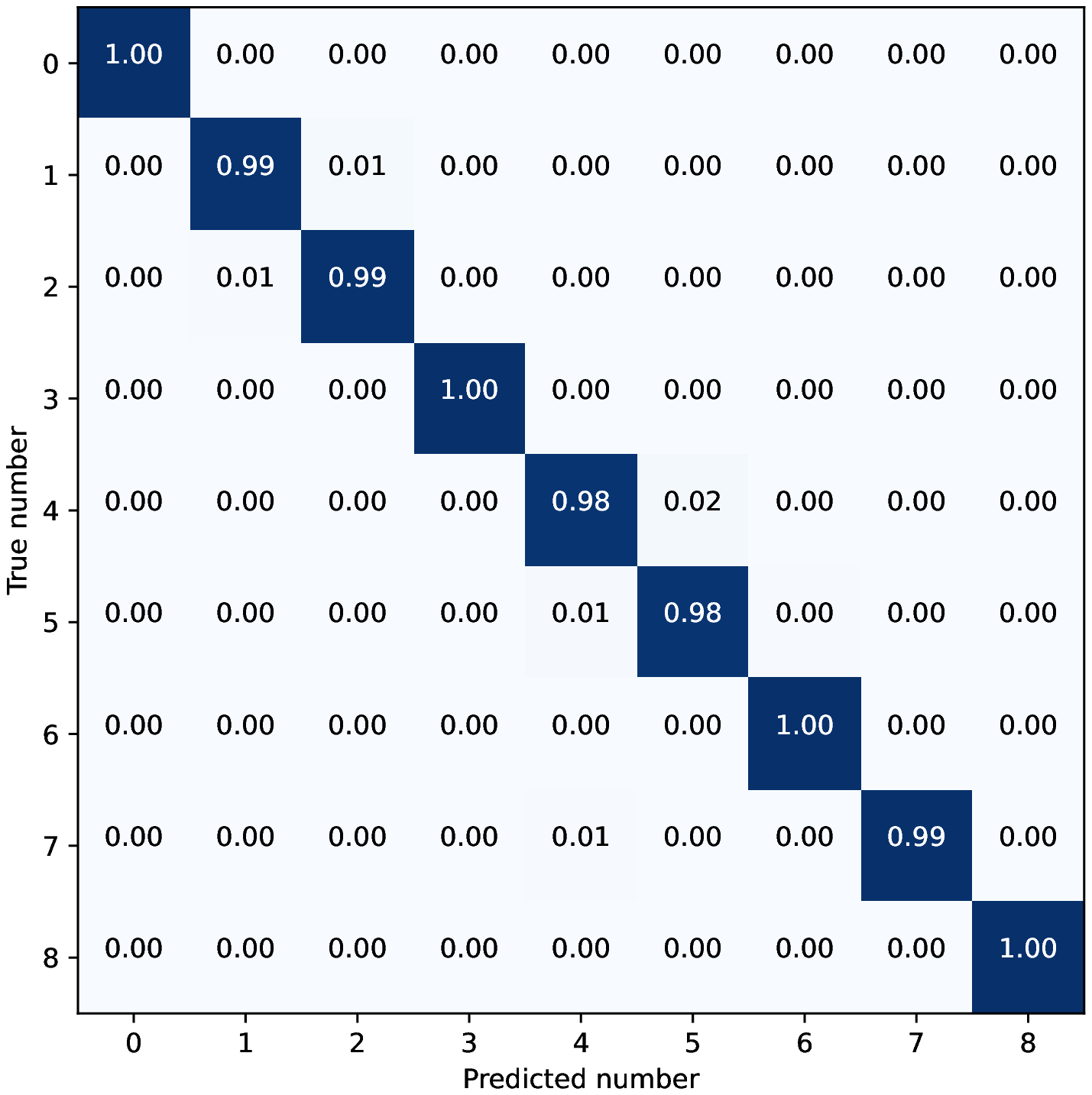}}
     \hfil
     \subfloat[Confusion Matrix of Mixed type]{
     	\label{Fig.sub.25}
     	\includegraphics[width=0.30\textwidth]{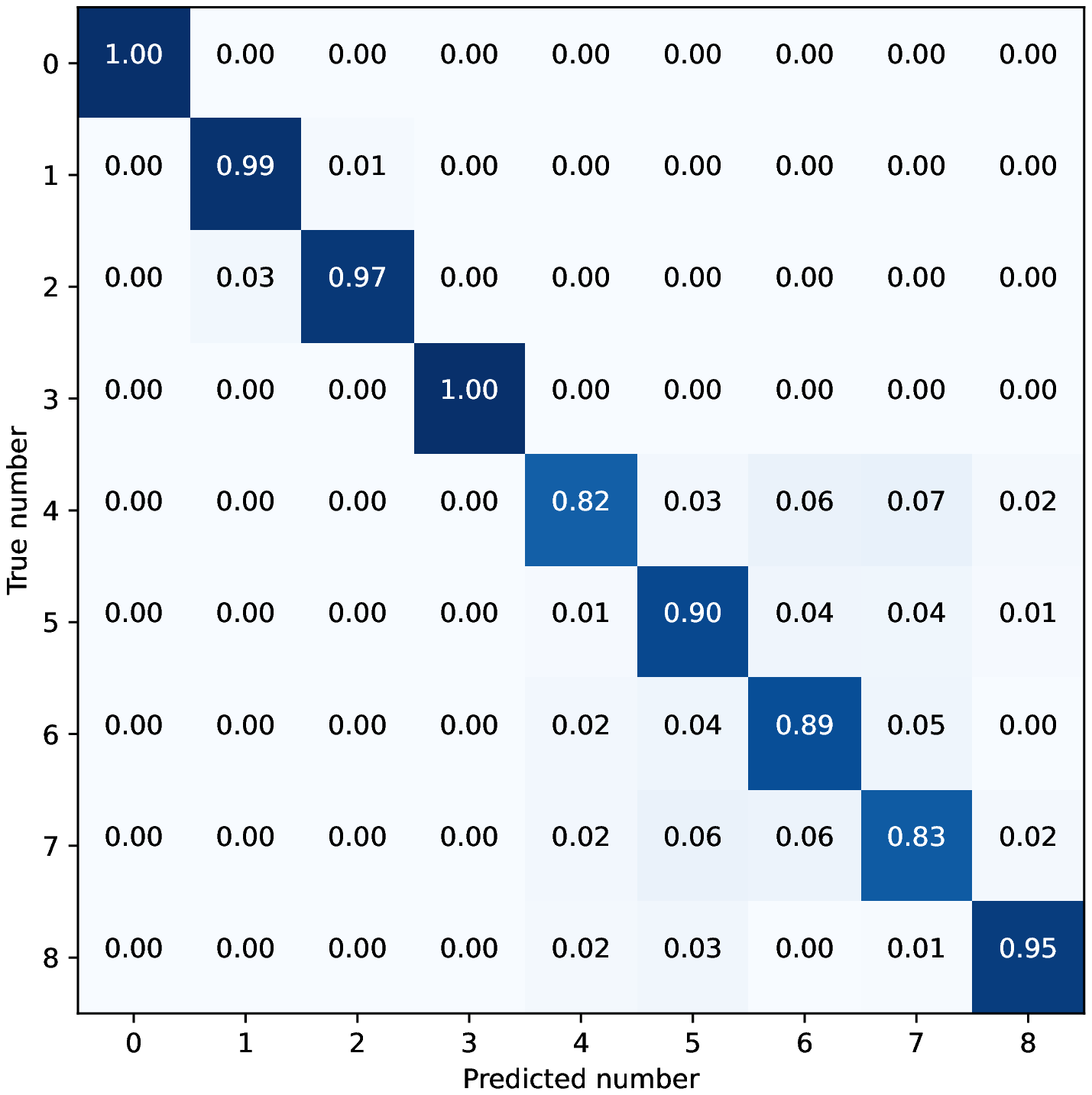}}
	\caption{Confusion matrix of classification results in the Room B NLOS scenario.}
	\label{fig_cm}
\end{figure*}

Fig. \ref{fig_cm} shows the confusion matrixes of ICC results in the Room B NLOS scenario. The $(i,j)$th element denotes the probability that the ground-truth $i$ people is identified as $j$ people.
We see that the detection accuracy decreases slightly when more people participate in the experiment. Overall, DASECount achieves very high accuracy (i.e.,over 99\%) where the error margins are mostly within 1-2 people. 
If we consider a presence detection problem, i.e., detecting whether there are any people in the room, with 5-shot learning, the proposed DASECount method achieves 100\% accuracy. 

\begin{figure}[H]
	\centering
	\includegraphics[width=0.5\textwidth]{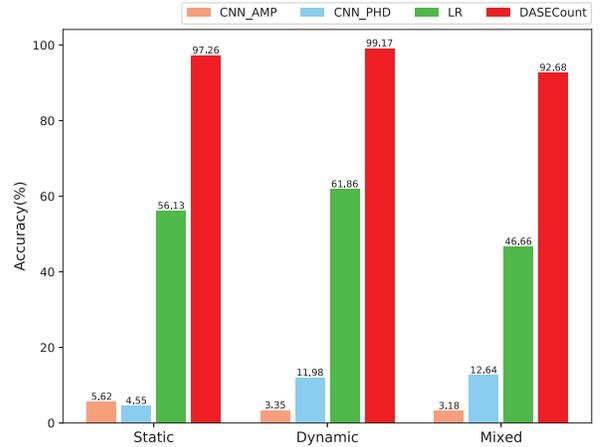}
	\caption{Detection accuracy comparisons of different cross-domain ICC methods.  \textit{CNN\_AMP} and \textit{CNN\_PHD} represent the CNN amplitude and phase difference feature extractor submodels, respectively. \textit{LR} is a logistic regression classifier directly trained with 5 samples. \textit{DASECount} is the proposed method.}
	\label{fig_6}
\end{figure}

\subsubsection{Advantage of cross-domain learning}
To evaluate the performance of the proposed DASECount method, we present the classification accuracy when the following three benchmark classifiers are used in the Room B NLOS scenario:
\begin{itemize}
	\item [(1)] 
	Source-domain CNN amplitude feature extractor;
	\item [(2)] 
	Source-domain CNN phase difference feature extractor;
    \item [(3)] 
    Directly train the LR classifier based on the 5 samples;
%    \item [(4)] 
%    The proposed method: First, input 5 samples of the target domain into CNN to extract features, then input the extracted features into the LR classifier for training, and finally use the trained classifier for prediction.
\end{itemize}

Fig. \ref{fig_6} shows the comparison result. If we directly apply the CNN feature extractor to the target domain ICC task without FSL, the accuracy is very poor, this verifies our claim that environment difference has a large impact on the ICC performance. The LR classifier is trained with raw CSI data (flattening CSI data to a vector without feature extractor processing). It achieves an average accuracy of 54.88\% in the scenario. This is because the few target-domain data samples are not sufficient to train a classifier with high-dimension of raw input data. Overall, the proposed DASECount framework outperforms the benchmark methods by at least 30\% in all the motion types considered. 

\begin{table}[htbp]
	\centering
	\caption{Influence of Combined Features}
	\begin{tabular}{ c | c c c c }
		\toprule
		&           & AMP & PHD & DASECount \\
		\midrule
		\multirow{3}{*}{1 shot} & Static    & 81.15\%  & 74.79\%  & \textbf{85.64}\% \\
		& Dynamic   & 91.45\%  & 76.28\%  & \textbf{93.29}\% \\
		& Mixed     & 67.56\%  & 59.04\%  & \textbf{76.26}\% \\
		\midrule
		\multirow{3}{*}{5 shot} & Static    & 90.89\%  & 91.82\%  & \textbf{97.26}\% \\
		& Dynamic   & 95.05\%  & 91.50\%  & \textbf{99.17}\% \\
		& Mixed     & 88.20\%  & 84.91\%  & \textbf{92.68}\% \\
		\bottomrule
	\end{tabular}%
	\label{tab:amp_phd}%
\end{table}

\subsubsection{Advantage of combined amplitude and phase features} \label{combine}
DASECount combines features in CSI amplitude and phase difference information from the target domain to train a lightweight LR classifier. 
Since the feature extractor contains independent amplitude and phase difference submodels, the target domain LR classifier can be trained using features extracted from one of the submodels alone.
Specifically, the vector length of the joint feature (1 $\times$ 1152) is twice as long as that of the single amplitude or phase feature (1 $\times$ 576).
Table \ref{tab:amp_phd} shows the comparison results of LR classifiers in the Room B NLOS scenario, where the \textit{AMP} represents the LR classifier is trained only with the amplitude feature and the \textit{PHD} represents the LR classifier is trained only with the phase difference feature. 
Compared with amplitude or phase difference feature alone, the accuracy improvement of the LR classifier trained on joint features ranges from the lowest 1.84\% (Dynamic type) to the highest 17.22\% (Mixed type) for 1-shot learning. For 5-shot learning, the accuracy improvement ranges from the lowest 4.12\% (Dynamic type) to the highest 7.77\% (Mixed type).

\subsubsection{Selection of feature map} \label{FE_Select}
The CNN feature extractor of DASECount has 6 convolutional blocks and a fully connected layer, and the extracted features of each layer have different dimensions and contain different information. 
We have tested the features from the last two convolutional blocks and the fully connected layer to train the LR classifier. 
Table \ref{tab:feature} shows the performance of LR classifiers trained with 3 different kinds of features in the Room B NLOS scenario. 
Compared with features from the fully connected layer (\textit{FC} in the table) and the final convolutional block (\textit{CNN-1} in the table), features from the penultimate convolutional block result in the best ICC accuracy in all cases.

\begin{table}[htbp]
	\centering
	\caption{Influence of feature map from CNN}
	\begin{tabular}{ c | c c c c }
		\toprule
		                        &  & FC(1$\times$18)  & CNN-1(1$\times$128)  & CNN-2(1$\times$1152) \\
		\midrule
		\multirow{3}{*}{1 shot} & Static    & 61.01\%  & 66.18\%  & \textbf{85.64}\% \\
	                    	    & Dynamic   & 71.01\%  & 84.41\%  & \textbf{93.29}\% \\
		                        & Mixed     & 39.03\%  & 60.27\%  & \textbf{76.26}\% \\
		\midrule
		\multirow{3}{*}{5 shot} & Static    & 83.92\%  & 90.14\%  & \textbf{97.26}\% \\
		                        & Dynamic   & 87.54\%  & 93.39\%  & \textbf{99.17}\% \\
		                        & Mixed     & 79.14\%  & 83.79\%  & \textbf{92.68}\% \\
		\bottomrule
	\end{tabular}%
	\label{tab:feature}%
\end{table}

\subsubsection{Selection of classifier model}
We compare different target domain classifiers: logistic regression (LR), support vector machine, and K-nearest neiber (NN) in the Room B NLOS scenario. Experiment results in Table \ref{tab:classifier} show the three machine learning classifiers have similar performance, while the LR classifier is slightly better than the other two by about 2\% detection accuracy.

\begin{table}[htbp]
	\centering
	\caption{Performance of different machine learning classifiers}
	\begin{tabular}{ c | c c c c }
		\toprule
	                    	&                & LR       & SVM      & NN \\
		\midrule
		\multirow{3}{*}{1 shot}  & Static    & \textbf{85.64}\%  & 79.93\%  & 79.60\% \\
		                         & Dynamic   & \textbf{93.29}\%  & 91.59\%  & 87.01\% \\
		                         & Mixed     & \textbf{76.26}\%  & 73.06\%  & 72.41\% \\
		\midrule
		\multirow{3}{*}{5 shot}  & Static    & \textbf{97.26}\%  & 96.07\%  & 95.12\% \\
	  	                         & Dynamic   & \textbf{99.17}\%  & 98.78\%  & 98.18\% \\
		                         & Mixed     & \textbf{92.68}\%  & 91.78\%  & 88.59\% \\
		\bottomrule
	\end{tabular}%
	\label{tab:classifier}%
\end{table}

\subsubsection{Effect of distillation} \label{effect of distillation}
\begin{figure}[H]%[!t]
	\centering
	\includegraphics[width=0.5\textwidth]{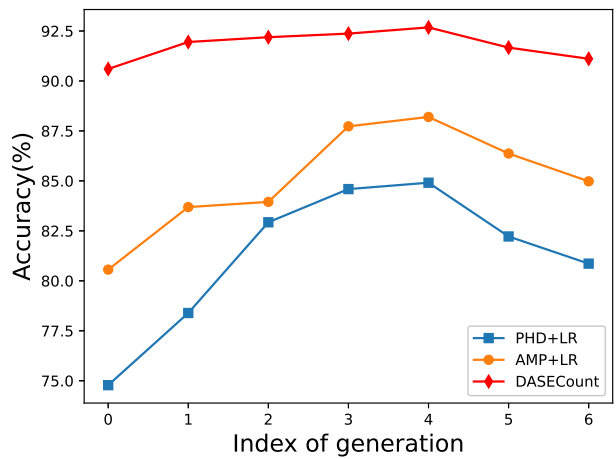}
	\caption{Influence of the generation of knowledge distillation on the performance of target domain classifier. \textit{PHD+LR} represents the phase difference submodel as the feature extractor. \textit{AMP+LR} represents the amplitude submodel as the feature extractor. \textit{DASECount} represents the proposed method, which combines both of amplitude and phase difference submodels as the feature extractor.  }
	\label{fig_distillation_result}
\end{figure}

In the meta-training stage, we have distilled the amplitude and phase difference submodels 6 times and obtained 7 generations of models in total. 
We conduct experiments under the Mixed type ICC in the Room B NLOS scenario and evaluate the performance of the target domain LR classifier using different generation models as the feature extractor. 
The influence of different distillation generation models is shown in Fig. \ref{fig_distillation_result}. 
As can be seen from the figure, compared with the model of generation 0, the accuracy of the target domain classifier can be improved by 3-10\% after several rounds of distillations, but the performance of the proposed DASECount drops after 4 rounds of distillation. Therefore, we need to track the ICC accuracy of each round of distillation and select the best one to produce the feature extractor. 

{\subsubsection{Compared with MAML FSL method}
Fig. \ref{fig_7} shows the comparison in scenario Room B NLOS with 5-shot learning. In particular, we compare the performance of the proposed FSL-based DASECount method with the well-known MAML method. It shows that the accuracy of the proposed DASECount method is on average 27.86\% higher than MAML when using only amplitude as the input measurement, and 13.6\% higher when using only amplitude as the input measurement. In both cases, the proposed method significantly outperforms the benchmark MAML. Besides, we can further improve the accuracy by combing the phase and amplitude features under the proposed DASECount framework.

\begin{figure}[H]
	\centering
	\includegraphics[width=0.5\textwidth]{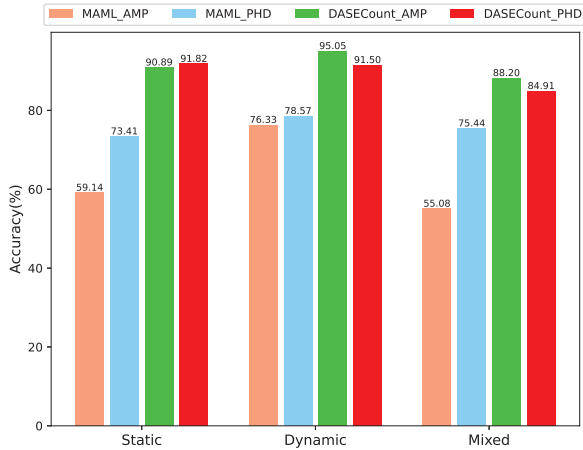}
	\caption{Detection accuracy comparisons of proposed DASECount and MAML in Room B NLOS scenario with 5-shot learning. }
	\label{fig_7}
\end{figure}

\section{Conclusions and Discussions}
In this paper, we have proposed a DASECount framework based on FSL to achieve highly accurate and robust cross-domain ICC performance. DASECount contains a meta-training stage and a meta-testing stage.  
In the meta-training stage, DASECount trains a CSI feature extractor consisting of the amplitude and phase difference CNN submodels with supervised learning. 
We also applied a knowledge distillation procedure to iteratively update the parameters of the CNN submodels for better generalization performance.  
In the meta-testing stage, thanks to the feature extractor that generates low-dimension features of the target domain data, DASECount attains very high cross-domain ICC accuracy with a simple lightweight LR classifier given very limited target domain data samples.  
Experimental results show that the proposed DASECount achieves over 92.68\%, and on average 96.37\% detection accuracy, in a 0-8 people counting task under various domain setups, which significantly outperforms the other representative benchmark methods considered. 
Overall, the proposed DASECount framework significantly enhances the robustness of cross-domain ICC tasks and reduces the operating cost in large-scale deployment of future WiFi-based indoor sensing applications. 

Notice that the training of feature extractor is performed offline just once on the source domain dataset. We can therefore perform the training on a powerful server using the sufficiently large source domain dataset, and fix the parameters of the feature extractor after the training converges. In our simulations, the complete CNN feature extractor contains 189513 training parameters and the training process converges in less than 5 minutes. 
In the target domain, we reuse the obtained feature extractor and only need to train a lightweight LR classifier consisting of only 1153 training parameters. Besides, the training is performed on a very small target domain data set following the FSL paradigm. Therefore, the classifier in the target domain can be quickly trained with the few-shot samples, and it takes less than 1 second in our simulations to complete the training. Overall, the proposed DASECount framework can be quickly extended to perform ICC in new target domains with very low computational complexity.

It is an important working direction for us to further improve the robustness of DASECount in some application cases. 
	For example, if the location of the WiFi transceiver or the background environment of devices changes significantly, it may cause sample data distribution shift and affect the training performance of the classifier. 
	In this case, it requires recollecting labeled data samples and retraining the classifier, which however is very costly if the change happens frequently.  
	A more feasible yet much more challenging solution is for DASECount to adapt to new domains with an unsupervised learning method. A promising method is to design a zero-shot framework based on a generative adversarial network (GAN), which requires no labeled data at all.  
	This is considered as an important future work.

%To further improve the robustness of DASECount in 
%Such as the adjustment of WiFi transceiver and the 
%
%In some application cases, such as the location of WiFi transceiver is evidently adjusted   directly deploying an ICC system to a new scenario, which is , it is difficult to obtain even a small number of labels samples. 
%Therefore, it is a main work direction for us to solve cross-domain ICC problem with zero-shot learning, i.e., training target domain classifier without labeled samples in the future. Related zero-shot works have been discussed in computer vision and image recognition fields. It is possible to carry out zero-shot learning on data samples with similar distribution.
%Besides, because the application scenarios of ICC are diverse and complex, testing model performance in a larger crowd range with more diverse target scenarios is also an important work direction for us to improve the architecture of DASECount. 
%In addition, DASECount can be deployed on cloud-edge computing platforms to conduct a real-time ICC system.

\bibliographystyle{IEEEtran}
\bibliography{reference}

\end{document}